\documentclass[runningheads]{llncs}
\renewcommand{\arraystretch}{1.0}

\usepackage[T1]{fontenc}
\usepackage{graphicx}
\usepackage{paralist}
\usepackage{tabularx}
\usepackage{comment}
\usepackage{enumitem}
\usepackage{comment}
\usepackage{subcaption}
\usepackage{wrapfig}
\usepackage{threeparttable}
\usepackage{makecell}
\usepackage{etex}
\usepackage{hyperref}
\usepackage{diagbox}
\usepackage[utf8]{inputenc}
\usepackage[T1]{fontenc}
\usepackage[ngerman,english]{babel}
\usepackage{times}
\usepackage{multirow}
\usepackage[dvipsnames,table,x11names]{xcolor}
\usepackage{booktabs}
\usepackage{paralist}
\usepackage{amssymb, amsfonts}
\usepackage{amsmath}
\usepackage{amsopn}
\usepackage{url}
\usepackage{multirow}
\usepackage{relsize}
\usepackage{xfrac}
\usepackage{pdflscape}
\usepackage{cite}
\usepackage{tabularx}
\usepackage{mdframed}
\usepackage{marginnote}
\usepackage{algorithm}
\usepackage{algorithmicx}
\usepackage[noend]{algpseudocode}
\usepackage{inconsolata}
\usepackage{nicefrac}
\usepackage{xspace}
\usepackage{tabu}
\usepackage{todonotes}
\usepackage{listings}
\usepackage{array} 
\usepackage{ragged2e}

\hypersetup{
	pdftoolbar=true,        % show Acrobat?s toolbar?
	pdfmenubar=true,        % show Acrobat?s menu?
	pdffitwindow=false,     % window fit to page when opened
	pdfstartview={FitH},    % fits the width of the page to the window
	pdftitle={},    % title
	pdfauthor={anonym},     % author
	pdfsubject={},   % subject of the document
	pdfcreator={},   % creator of the document
	pdfproducer={}, % producer of the document
	pdfkeywords={}, % list of keywords
	pdfnewwindow=true,      % links in new window
	colorlinks=true,       % false: boxed links; true: colored links
	linkcolor=Brown,          % color of internal links (change box color
	citecolor=Brown,        % color of links to bibliography
	filecolor=Brown,      % color of file links
	urlcolor=Brown           % color of external links
}

\newcommand{\mypar}[1]{\noindent\textbf{#1.}}

\newcommand{\update}[3][0em]{\marginnote{}[#1]{#3}}
\begin{document}
\title{A Procedural Framework for Assessing the Desirability of Process Deviations}
\titlerunning{Framework for Assessing Deviation Desirability}

\author{Michael Grohs\inst{1}\orcidID{0000-0003-2658-8992} \and
Nadine Cordes\inst{1,2}\orcidID{0009-0003-9547-7919} \and
Jana-Rebecca Rehse\inst{1}\orcidID{0000-0001-5707-6944}}
\authorrunning{M. Grohs et al.}
\institute{University of Mannheim, Germany \\
\email{\{michael.grohs@,nadine.cordes@students.,rehse@\}uni-mannheim.de}
\\
 \and
SAP Signavio, Berlin, Germany
}

\maketitle              % typeset the header of the contribution
\begin{abstract}
%The abstract should briefly summarize the contents of the paper in 150--250 words.

Conformance checking techniques help process analysts to identify where and how process executions deviate from a process model. 
However, they cannot determine the desirability of these deviations, i.e., whether they are problematic, acceptable or even beneficial for the process. 
Such desirability assessments are crucial to derive actions, but process analysts typically conduct them in a manual, ad-hoc way, which can be time-consuming, subjective, and irreplicable.
To address this problem, this paper presents a procedural framework to guide process analysts in systematically assessing deviation desirability. It provides a step-by-step approach for identifying which input factors to consider in what order to categorize deviations into mutually exclusive desirability categories, each linked to action recommendations. 
The framework is based on a review and conceptualization of existing literature on deviation desirability, which is complemented by empirical insights from interviews with process analysis practitioners and researchers. 
We evaluate the framework through a desirability assessment task conducted with practitioners, indicating that the framework effectively enables them to streamline the assessment for a thorough yet concise evaluation.

\end{abstract}

\section{Introduction}

To ensure organizational success, it is essential to understand how business processes align with their intended behavior \cite{CarmonaCCfund}. 
This behavior is typically captured by normative process models \cite{Dunzer_2019SOA_CC}, which can be automatically compared to actual process execution traces by means of conformance checking \cite{laghmouch2020classifying}. 
Conformance checking techniques identify whether and how those traces differ from the behavior specified in the model \cite{Dunzer_2019SOA_CC}. 
Any differing part of the traces, such as skipped or unnecessarily repeated activities, indicates a deviation between the trace and the model \cite{CarmonaCCfund}.

Although conformance checking can identify where and how deviations occur, it treats all deviations equally. 
However, for addressing deviations, it is crucial to consider their desirability, i.e., whether they are problematic, acceptable, or even beneficial~\cite{depaire2013process}. 
Many deviations are false alarms that do not require any action \cite{yang2018approach}. 
One example is a logging error, where the recorded behavior is impossible, like a \textit{logout} without prior \textit{login} \cite{adamo2018_business}. 
Other deviations can represent desired behavior, e.g., a streamlined approach that saves time \cite{Dumas2015enabling} or an innovative approach that increases customer satisfaction \cite{alter2014theory}. 

Accordingly, process analysts need to assess and categorize the desirability of process deviations to derive actionable insights. 
Since conformance checking techniques do not support this, analysts are left to conduct their own desirability assessments \cite{laghmouch2020classifying}. However, these assessments are only scarcely supported by literature \cite{depaire2024process}. Some studies propose desirability categories, but they lack structured approaches for how to reach them \cite{depaire2013process,alter2014theory,koschmider2021demystifying,dunzer2024conceptualizing}. Moreover, these categories often overlap conceptually or focus on different aspects, such as compliance versus performance \cite{andrade2016factors}, leading to redundant or scattered insights. As a result, analysts rely on manual, ad-hoc assessments that lack consistency, replicability, and transparency. This makes it difficult to unify insights across analysts and impedes well-informed decision-making about reactive measures.

To address this problem, this paper introduces a procedural framework for assessing the desirability of process deviations. This framework provides step-by-step guidance for analysts to assess the desirability of past deviations and derive actionable insights for proactively managing future deviation. 
The framework is built on existing studies on deviation desirability, identified in a literature review, 
as well as empirical insights from interviews with process analysis practitioners and researchers. 
%to close this gap. 
This two-pronged approach allows us to define which input factors should be considered in what order to categorize deviations into mutually exclusive desirability categories, for which actions can be recommended. 
The resulting framework comprises five assessment steps, which are arranged to reduce analysis effort by employing input factors as knockout criteria, allowing the early determination of certain desirability output categories. 
\update{R1.4}{Thus, the framework has two properties: its components are correct and complete (P1) and it is useful for process analysts (P2).}
Our evaluation results illustrate this as process analysts are able to effectively apply the framework for a comprehensive categorization of concrete deviations, benefiting from the offered orientation and streamlined analysis. 

The remainder of the paper is organized as follows: After providing an overview of related work in Sect.~\ref{sec:rw}, we outline our research method for framework development in Sect.~\ref{sec:method}. The resulting framework is presented in Sect.~\ref{sec:results} and evaluated in Sect.~\ref{sec:eva}. Finally, we discuss our findings and conclude the paper in Sect.~\ref{sec:conlusion}.

\section{Related Work}\label{sec:rw}

Deviation desirability has been analyzed from different angles. 
A common distinction is made between \textit{exceptions} and \textit{anomalies} \cite{laghmouch2020classifying,depaire2013process,swinnen2012process}, where former have neutral or positive and latter have negative impact on the process \cite{depaire2013process}. 
Some works differentiate between explicit and implicit exceptions as well as intended and unintended anomalies \cite{depaire2013process,swinnen2012process,van2017behavioral} and consider the importance of context \cite{park22context}.
%To assess whether deviations are exceptions or anomalies, one approach classifies traces into these two categories \cite{laghmouch2020classifying}.
Similarly, deviations are often subdivided into true and false alarms \cite{yang2018approach} or positive and negative deviations \cite{Dumas2015enabling}. This distinction is especially relevant for auditing, where auditors must find problems in sets of deviations \cite{jans2019how,li2016exception,laghmouch2020classifying}. \update{M1, R3.1}{Although these works acknowledge that deviations can be non-problematic, they do not provide guidance on how to assess deviation desirability.}

Other research extends deviation desirability to so-called \textit{workarounds}, i.e., goal-oriented and intended diversions from prescribed behavior by humans\cite{alter2014theory}. 
They can be exceptions or anomalies and can be detected semi~\cite{waal2024sword} or fully automatically~\cite{weinzierl2022detecting}. \update{M1, R3.1}{Workarounds only focus on intended behavior, whereas deviations can be incidental.}

The desirability of deviations can also be quantified by assigning a severity score to the non-conformance of certain activities.
This score can be expressed as a criticality ranking, combining severity and likelihood of deviation-associated risks \cite{alizadeh2016risk}. 
When this score is defined during the design of normative process models, it can be used in a cost function to quantify the degree of non-conformance \cite{stertz2020data,adriansyah2013controlling}, also including deviations beyond the control-flow perspective \cite{zhang2020towards}. 
In this manner, we can find so-called ``break-the-glass'' deviations, which are conceptually similar to workarounds \cite{adriansyah2013controlling}. 
\update{M1, R3.1}{The severity score however must be defined a priori and only considers undesired behavior.}

Risk-aware BPM makes use of ontological dependencies of deviations, e.g., causal activity relationships they violate \cite{andree2023beyond}, to get an idea of their severity and potential consequences \cite{andree2024amIallowed}.
This violation perspective distinguishes between noise (logging errors), reflecting ontologically impossible behavior, and outliers (anomalies or exceptions) \cite{koschmider2021demystifying}. Similarly, declarative constraint violations can be categorized as violations of laws of nature (suggesting logging errors), goal violations (indicating reduced process quality), or norm violations (signaling compliance issues) \cite{adamo2018_business}. 
Such violations were conceptualized w.r.t. their relation to workarounds, anomalies, or fraud \cite{dunzer2024conceptualizing}. \update{M1, R3.1}{This allows to characterize violations into descriptive categories, but does not imply clear assessment steps or action recommendations. 
Also, only the negative impact of deviations is depicted.} 

In summary, there are several views on deviation desirability. Some have a similar understanding of desirability categories, whereas others differ considerably. The connection between those views on anomalies, exceptions, workarounds, ontological dependencies, and deviation severity remains unclear. 
For negative deviations, steps towards a unified view spanning several concepts have been taken \cite{dunzer2024conceptualizing}, \update{M1, R3.1}{yet such an overview for the full spectrum of deviation desirability remains missing, especially w.r.t. guidance on the desirability assessment itself.}
%Further, there is a lack of action-oriented, structured guidance for how to classify a deviation based on its desirability.  
%First steps towards a unified view on negative deviations have been made \cite{dunzer2024}, an action-oriented procedural guide for ..
In this paper, we integrate the different concepts into one framework with actionable, mutually exclusive desirability categories and provide a procedural structure for deviation categorization.
%In this paper, we summarize these concepts into one framework with mutually exclusive deviation categories that go beyond two categories and provide a holistic understanding of the desirability. 

\section{Research Method}\label{sec:method}

\begin{figure}[ht]
  \begin{center}
    \includegraphics[trim=0 280 5 0,clip,width=0.98\textwidth]{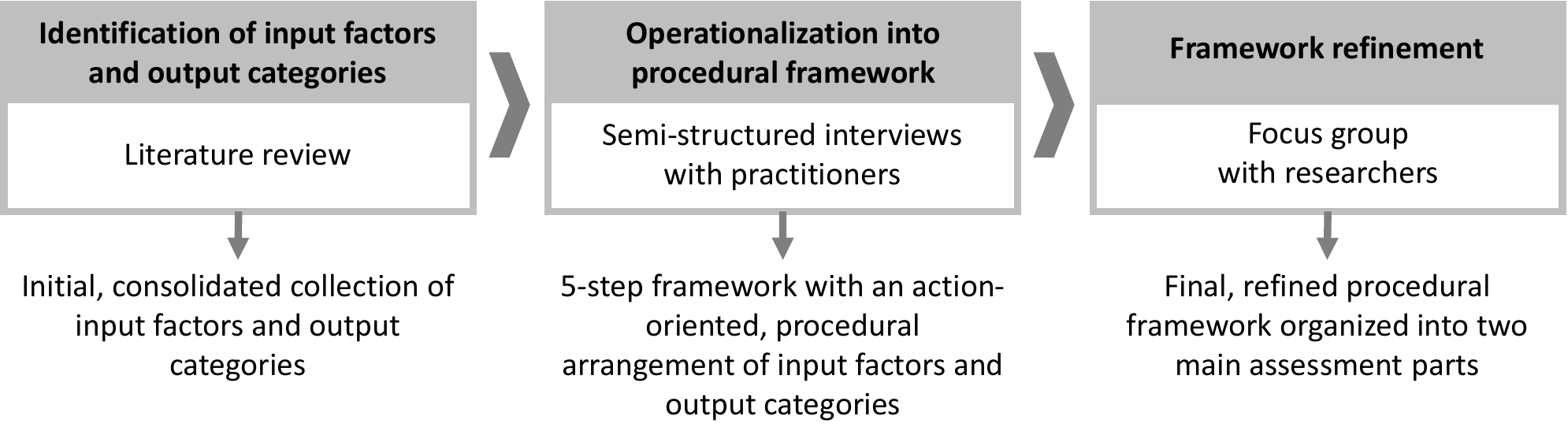}
  \end{center}
  \vspace{-1.5em}
  \caption{Overview of framework development phases}
  \label{fig:Method_phases}
  \vspace{-1.5em}
\end{figure}

To develop a framework for deviation desirability \update{R1.4}{that is correct and complete (P1) as well as useful (P2)}, we used a three-phase research method, shown in Fig.~\ref{fig:Method_phases}. 
First, we collected input factors and output categories from existing desirability assessments in the literature  to be used as the basis of our framework. Second, we organized these elements into assessment steps to create the action-oriented procedural structure of our framework. To this end, we held semi-structured interviews with process analysis practitioners designed as a deviation desirability assessment task. Last, we conducted a focus group with researchers to refine the framework and improve its conceptual correctness and clarity. We explain these phases in the following.

\subsection{Identification of Input Factors \& Output Categories}
In the first phase, we aimed to identify input factors and output categories typically used to assess deviation desirability to serve as the basis of our framework. 
To this end, we conducted a literature review according to vom Brocke et al.~\cite{brocke2015standing}. 

\mypar{Search Terms} An initial scan showed two main types of relevant sources: conceptual work categorizing deviations and empirical work assessing real-life deviations. 
Hence, we formulated two search strings. Both contained \textit{``business process''} to root our search in the business process management (BPM) field and \textit{(devia* OR conformance OR compliance OR anomaly OR exception OR workaround)} to include all terms often used to describe deviations. 
Then, we added \textit{``case study''} for the search on empirical works and \textit{(framework OR categor* OR taxonomy OR classif*)} for conceptual papers. 
To include papers that discuss the general desirability of deviations, we formulated a third search string with \textit{(desirab* OR severity OR impact OR outcome)}.

\mypar{Databases} We chose Scopus, Academic Search Ultimate and Business Source Ultimate via EBSCOhost, AIS eLibrary, and IEEE Xplore as databases covering the fields of information systems and BPM. Considering all English papers from our search results led to an initial set of 5,196 papers, which we scanned for a thematic fit in title, abstract and keywords, stopping the search in each database if the last 100 results did not yield a fit. After removing duplicates, 23 fitting papers remained. We added another three studies that were known to be relevant from our previous research and incorporated six papers found through a forward-backward search, totaling 32 papers.\update{R2.6} {\footnote{Details on the paper selection process and a complete overview of all input factors and output categories identified through their review can be found online: \url{https://doi.org/10.6084/m9.figshare.28546982}}}

\mypar{Coding} We coded these 32 papers in two cycles to collect and consolidate the presented input factors and output categories, adopting the qualitative coding guidelines by Saldaña~\cite{saldana2015coding}. 
From the descriptions of what aspects were considered in a desirability assessment, we defined 23 input factors.   
Among these, many were tangible, e.g., \textit{data quality} \cite{koschmider2021demystifying}, \textit{process model adequacy} \cite{yang2018approach}, \textit{compliance} \cite{Dumas2015enabling,andree2024amIallowed,maggi2014predictive} and \textit{security}, but others were soft, e.g. \textit{intent} \cite{andrade2016factors,alter2014theory,outmazgin2023fromautomatic}
or \textit{management awareness} \cite{depaire2024process,laghmouch2020classifying,alter2014theory,outmazgin2013exploring}.
For the output categories, we collected 63 partially overlapping or synonymous ones at different levels of detail, which we organized into 12 thematic categories for a first consolidation. These included, e.g., \textit{logging error} \cite{yang2018approach}, \textit{outlier} \cite{koschmider2021demystifying}, or \textit{contextual deviation} \cite{koschmider2021demystifying,andree2024amIallowed,park22context}.

Although all identified input factors were stated to influence deviation desirability, this influence was often ambiguous, meaning that whether they render a deviation desirable often depends on the specific situation. As a result, the factors and any output categories defined on their basis lacked clear desirability implication. %Furthermore, many output categories in the papers were not only linked to desirability assessments, but mixed with distinctions of root causes of deviations. 
%Similarly, the output categories were often overlapping, thus not mutually exclusive and dependent on each other. 
Hence, in the next phase, we aimed to remove ambiguity and establish the procedural nature of our framework by complementing our findings with empirical insights.

\subsection{Operationalization into Procedural Framework}
In the second phase, we pursued a two-fold objective of operationalizing our findings. First, we wanted to ensure that each input factor and output category contributed to desirability assessments. %and remove any remaining ambiguity. 
Second, we sought to identify connections between them to arrange them in a procedural manner to reflect typical desirability assessment steps.
%milestones in such a desirability assessment to which we can link input factors and output categories so that we can develop the step-by-step structure that we envisioned for our framework.
To achieve this, we conducted six semi-structured interviews with process analysis practitioners, following the methodology by Myers and Newman~\cite{myers2007interview}. The interviews involved a deviation desirability assessment task centered on a fictional procurement process, allowing us to observe how process analysts approach such assessments.\footnote{Participant details, the full interview guide and all changes to the input factors and output categories are provided online: \url{https://doi.org/10.6084/m9.figshare.28546967}} 
%This process was chosen because it is relevant to most organizations and therefore familiar to many process analysts.

\mypar{Participants} 
Our participants were practitioners with solid experience in process analysis (on avg. 13 years) and in the procurement domain (on avg. 10.7 years). They either worked with process analysis internally, in a tool development role, or as consultants.
%, we otherwise opted for heterogeneity of experience with respect to company size and industries. 
%Since all participants conducted process analysis for companies of various size and industries, we further benefited from their aggregated experiences. 

\mypar{Questions} 
We created four deviation examples for a procurement process and asked the experts to think out loud while assessing their desirability. 
Subsequently, we introduced the input factors from the previous phase to see how they might alter the experts' desirability judgments, aiming to link specific input factors and output categories. Also, we inquired about actions they would take based on selected output categories to ensure these categories support an action-oriented, proactive management of future deviations.

\mypar{Coding} 
All interviews were transcribed and coded using Mayring's template approach for qualitative content analysis~\cite{mayring2014coding}. Based on that, we refined the input factors by merging them into high-level concepts where experts indicated that their distinction was irrelevant for deviation desirability. For instance, they suggested integrating \textit{security} with \textit{compliance}, as security is often reflected in internal policies or laws. We also differentiated factors that impact desirability from those that merely indicate desirability likelihood. For example, a positive \textit{intent} might suggest desirability but does not guarantee it, especially if process participants cannot oversee the consequences of their actions. Since these indicators are not meaningful for the desirability assessment, we excluded them from our framework, which eventually reduced the input factors to 15.

For the output categories, we maintained some from our literature coding after minor adjustments, such as renaming \textit{logging errors} and \textit{model issues} to \textit{false alarm (log)} and \textit{false alarm (model)} to emphasize the similarity in their implications. 
%However, many other output categories were subcategories referring to details, such as security, that we had already identified as irrelevant for desirability. 
Other output categories referred to factors like security that we had already removed
from our framework.  
Hence, we excluded them and opted for universally applicable and actionable categories: \textit{positive, neutral,} and \textit{negative}. This led to a total of seven output categories.

After that, we conducted a second coding cycle to develop the framework's procedural structure. We searched for patterns in the experts' descriptions to identify key milestones in their approach of assessing the presented deviations. 
Specifically, we marked a critical milestone if we detected combinations of input factors that served as knockout criteria for the early identification of certain output categories. We then translated these milestones into a step-by-step structure of five steps, to which input factors and output categories are systematically mapped.  
%For example, the experts used the factor \textit{data quality} to rule out \textit{false alarms} in form of logging errors and thus terminate the analysis right after the first step.
Finally, we linked each output category to the an action recommendation provided by the experts: \textit{filter out, ignore, prevent}, and \textit{adopt}.
%where we used the wording from \cite{beerepoot2018prevent} for the last three. 
The result was a first version of our framework, which we refined in the next phase.
%The resulting framework requires a process deviation and its details for the corresponding input factors and returns a desirability evaluation as one of X output categories of the framework. 

\subsection{Framework Refinement}
To improve the conceptual correctness and clarity of the first framework version, we continued with a dedicated refinement phase. We followed a focus group approach as outlined by Krueger and Casey \cite{krueger2014focus}, which fosters in-depth discussion and thus facilitates the collection of detailed qualitative feedback on all framework elements.\footnote{As for the interviews, the supplementary material contains an overview of participants, interview guideline and changes made: \url{https://doi.org/10.6084/m9.figshare.28546973}}  

\mypar{Participants} This time, we chose a conceptual focus to reflect the nature of the refinement goal and thus invited three researchers to our focus group. They all had several years of process analysis experience (on average 9 years), particularly in the analysis and interpretation part connected to managerial implication of conformance checking.  

\mypar{Questions} Again, we constructed a semi-structured guideline. 
After outlining the high-level framework structure, the five steps and their corresponding elements were presented in detail. We encouraged feedback on the overall structure of the framework, the correctness and clarity of the input factor and output category definitions as well as any other aspect that came to the experts' attention.  

\mypar{Coding} Similar to the interviews in the second development phase, the focus group was transcribed and coded using a template approach.
Based on the feedback of the focus group, we refined the framework in several ways.  
We partially adjusted the wording to increase expressiveness and performed a last input factor refinement in form of splits and merges, leading to a final number of eleven input factors. %especially for process model related factors.
Furthermore, the experts' feedback led to two re-arrangements in the framework. First, we incorporated a risk and opportunity perspective to not only account for known impacts from past deviations but also for unseen or less systematically occurring scenarios. Second, we split the framework into two parts: an individual-level analysis to filter out business-irrelevant deviations and an aggregated analysis to assess the collective, process-wide impact of remaining deviations across a set of deviating traces with similar deviation behavior.

This led to the final version of the framework with five steps across the individual level and aggregated level analysis, eleven input factors and seven output categories which we present in detail in the following section and evaluate in Sec.~\ref{sec:eva}.

\section{Results}\label{sec:results}
\begin{figure}[ht]
  \begin{center}
    \includegraphics[trim=11 455 17 0,clip,width=0.98\textwidth]{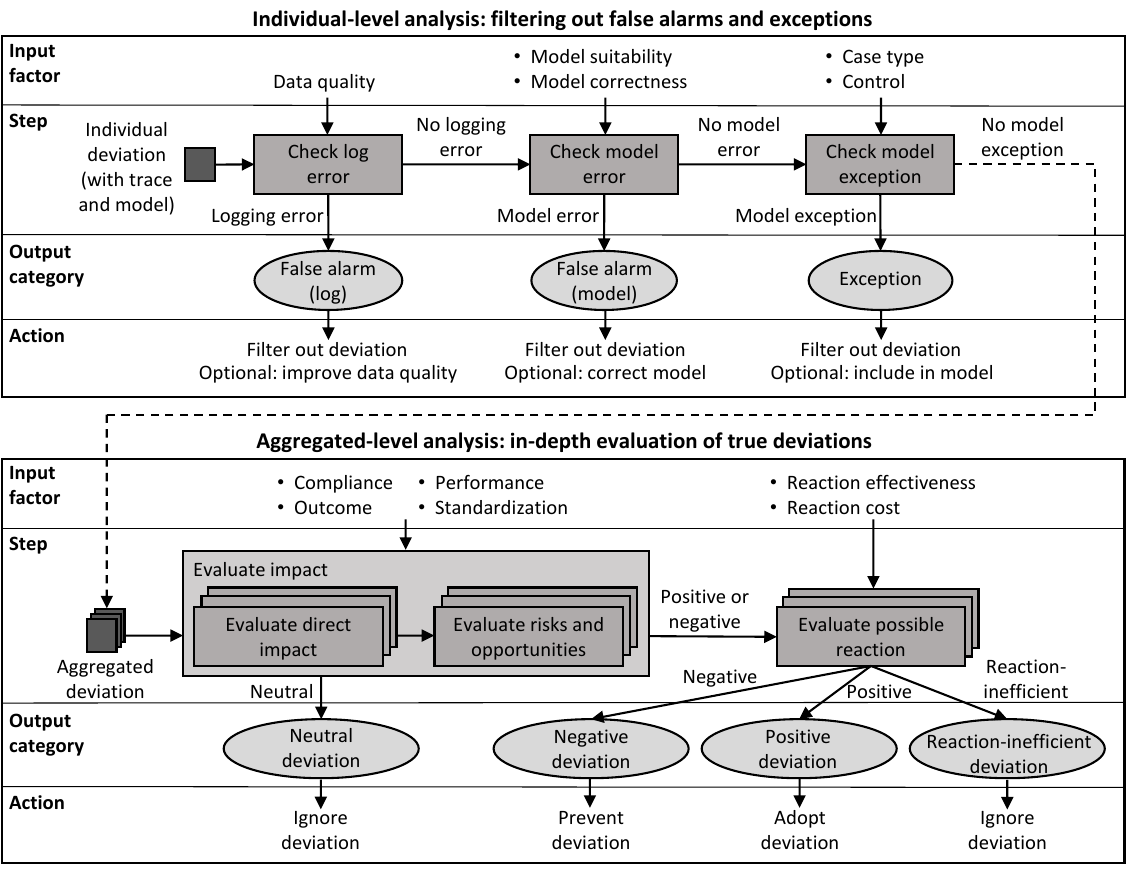}
  \end{center}
  \caption{Procedural framework for deviation desirability assessment}
  \label{fig:Framework}
  \vspace{-1em}
\end{figure}

This section presents our final framework for assessing the desirability of process deviations, illustrated in Fig.~\ref{fig:Framework}. This framework enables process analysts to systematically evaluate past process deviations identified through any conformance checking technique, with the aim of offering proactive recommendations to manage future deviations. These deviations are not limited to the control flow, but can also come from other perspectives such as resources, data, or time \cite{CarmonaCCfund}. To initiate the assessment, the framework requires both the trace with the detected deviation and the corresponding process model. Subsequently, it gathers and evaluates up to eleven input factors over the course of five assessment steps. Each step holds the possibility to already assign a final desirability output category on the basis of its input factors, which function as knockout criteria. Hence, subsequent steps are only taken if no conclusive output category has been assigned, reducing the overall assessment effort.

The framework classifies the desirability of the deviation into one of seven output categories, each linked to an action recommendation, in two main assessment parts:

\begin{compactenum}[(1)]
\item \textbf{Individual-level analysis:} The first part of the framework analyzes deviations on an individual level for the early identification and filtering out of deviations that are irrelevant from a business perspective. These deviations can be a \textit{false alarm (log)} resulting from poor data quality in the event log, a \textit{false alarm (model)} resulting from an error in the process model, or an \textit{exception} resulting from a justified business situation that was not included in the model for pragmatic reasons. If the deviation is neither of these, the framework renders it a true deviation, for which assessment continues in the second part.

\item \textbf{Aggregated-level analysis}: This part collectively assesses true deviations at an aggregated level. For this purpose, similar deviating behavior across multiple traces is aggregated and analyzed collectively. This serves as the basis for an insight into the process-wide desirability of deviations to derive actions geared towards future process improvements. Here, their direct and potential impact is analyzed to determine whether the deviations are \textit{positive}, \textit{negative}, or \textit{neutral}. Neutral deviations can be ignored, whereas positive and negative ones are evaluated further to determine whether reactive measures to them are worthwhile or if they are \textit{reaction-inefficient}.
\end{compactenum}

\begin{figure}[ht]
  \begin{center}
    \includegraphics[trim=0 330 3 0,clip,width=0.98\textwidth]{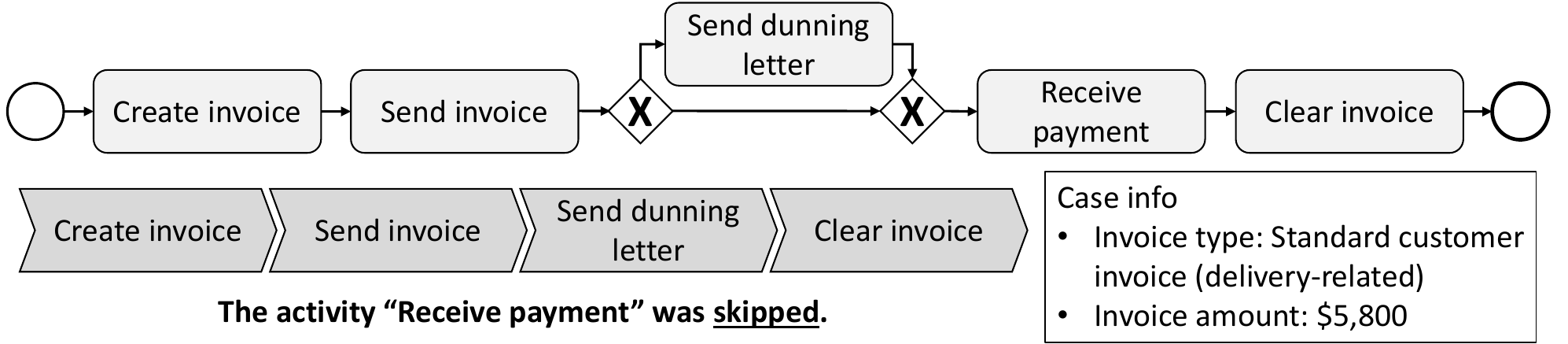}
  \end{center}
  \caption{Example deviation from the invoice-to-cash process}
  \label{fig:DeviationExample}
\end{figure}

\noindent
In the following, we introduce the two framework parts in detail. \update{M2, R2.3}{We illustrate it using the invoice to cash example from Fig.~\ref{fig:DeviationExample} with model, trace, deviation and case info.}

\subsection{Individual-level Analysis: Filtering Out False Alarms and Exceptions}
The first part of the framework analyzes deviations on an individual level to filter out deviations that arise from issues in the process representations used for conformance checking, i.e., the trace and the model. %, while the real process execution conforms to its intended behavior.
Such representational issues can lead to a flood of false alarms and exceptions \cite{yang2018approach,depaire2013process}, which are irrelevant from a business perspective. Thus, the framework checks for log errors, model errors, and model exceptions.

\mypar{Check log error} The first step involves assessing whether a deviation is caused by a logging error, which could indicate a false alarm. This check requires one input factor:

\begin{compactitem} 
\item \textit{Data Quality:} Insufficient data quality in the event log can result in traces that do not accurately depict what happened in reality \cite{koschmider2021demystifying}. 
\end{compactitem}

%\mypar{Decision} 
If the deviation can be clearly attributed to poor data quality of its trace%, i.e., the deviation did not really occur in the actual process execution
, it is labeled \textit{false alarm (log)}. %A prime example for this are violations of laws of nature such as a "login" only occurred after a "logout" \cite{adamo2018_business}. These violations can be rendered impossible by definition.
In these cases, the framework recommends to ignore the deviation from a business perspective, and thus filter it out since there was no problem in reality for a process analysts to act upon. From a technical perspective, there is the option to investigate whether data quality improvements would be helpful for future analyses. 
If no false alarm from the log is identified, the next step of the framework is taken.
\update{M2, R2.3}{In the example in Fig.~\ref{fig:DeviationExample}, there is no indication of bad data quality, thus the next step is entered.}
%log repair as for example explained in \cite{CarmonaCCfund} can be performed. After log repair, there is the option to repeat CC. This can help to ensure that no further issues that were initially masked by the logging error are missed. Accordingly, the classification with the decision tree could be repeated. 

\mypar{Check model error} The framework continues by assessing whether a deviation is caused by a misalignment between process model and process, potentially uncovering a false alarm caused by the model.
%. In that case, a deviation can indicate a false alarm since the trace might be desire but not reflected in the model. 
For this check, two input factors are required:

\begin{compactitem} 
\item \textit{Model correctness:} The process model might not reflect the desired behavior correctly, that is, there are mistakes in the model \cite{yang2018approach}. For example, a necessary activity could simply be forgotten in the model. 

\item \textit{Model suitability:} The process model might be generally unsuitable for the recor\-ded process. This can have various reasons, e.g., the model is outdated, a new model is used for an old event log, or the model depicts a different process.
\end{compactitem} 

%\mypar{Decision} 
If the deviation occurs only due to problems in the process model, it is labeled \textit{false alarm (model)}. 
%An appropriate model would avoid the deviation. 
In these cases, the recommendation of the framework is to filter out the deviation, and (optionally) to correct the model if necessary. If no false alarm is identified, the assessment continues with the third step.
\update{M2, R2.3}{The process model in Fig.~\ref{fig:DeviationExample} is suitable and correct, thus there is no false alarm and the next step is entered.}

\mypar{Check model exception}
Next, it should be assessed whether the deviation is a justifiable exception to the process model from a business point of view. Even if the
model is free of semantic errors and does not trigger any general false alarms, it may still have limitations in addressing specific exceptional business situations \cite{depaire2013process,laghmouch2020classifying}. Often, these limitations are intentional. Process models are crafted pragmatically, focusing on simplicity and readability, which means some exceptional situations are excluded, even if they are acceptable \cite{depaire2024process,CarmonaCCfund}. Therefore, to decide if a deviation qualifies as an exception, a detailed evaluation of the particular circumstances of the deviating case is essential. Two key factors should be taken into account:

\begin{compactitem} 
\item \textit{Case type:} Some cases fall outside the scope of business rules defined by the model due to the nature of the case itself \cite{maggi2014predictive}, referred to as the case type. If these cases do not conform to the model because it does not account for them separately, their behavior can still be justified. For example, a process model for handling invoices typically requires a link to a purchase order. However, this does not apply to cases involving tax payments, which do not have an associated purchase order.

\item \textit{Control:} Certain conditions within processes are beyond the control of the organization managing them \cite{park22context,maggi2020probablistic}. Deviations can represent necessary and justified responses to these conditions, although they are rarely included in process models, which usually focus on ``sunny day'' situations \cite{laghmouch2020classifying}. Generally, two conditions fall outside an organization's control: external contexts, such as regulations or supplier issues \cite{andree2024amIallowed,weinzierl2022detecting}, and past events that have led to the deviation as an unavoidable consequence \cite{adamo2018_business,andree2023beyond}. Although there is no control over these deviations in the short-term, organizations might re-gain control in the long-term, and thus the deviation might only be justified if no adequate reaction could have been taken yet.
\end{compactitem} 

%\mypar{Decision} 
If a deviation is not included in the model but is justified due to its case type or uncontrollable conditions, it is labeled as an \textit{exception}. In such cases, the framework recommends filtering out the deviation. %because it is appropriate for its specific context or nature. 
There is also an option to decide whether this behavior should be incorporated into the official model, keeping in mind the trade-off with maintaining model simplicity \cite{depaire2013process,CarmonaCCfund}.  If no exception is identified, the framework progresses into its second part to evaluate the deviation's impact in detail.
\update{M2, R2.3}{In the example in Fig.~\ref{fig:DeviationExample}, it is highly unlikely that skipping "Receive Payment" is an allowed case type or out of control since it is a central activity, thus the next phase is entered.}

\subsection{Aggregated-level Analysis: In-depth Evaluation of True Deviations}
The second part of the framework is entered only if a deviation is deemed a true deviation, meaning that it is not classified as a false alarm or an exception. True deviations should be assessed for their impact on the overall process and organization. This assessment should no longer view the deviation in isolation but consider the broader pattern of the behavior it represents. This provides insights beyond individual cases, which might otherwise be affected by random conditions, and allows for the inclusion of deviation frequency as a key aspect. 
To achieve this perspective, deviations should be aggregated into sets that are analyzed jointly in one of three ways:
\begin{compactitem}
\item a set of traces with the same deviating behavior in the same activity sequence
\item a set of traces with the same deviating behavior in differing activity sequences
\item a set of traces with 
%semantically 
similar deviating behavior in differing activity sequences 
\end{compactitem}
Thereby, similar deviations refer to situations where the behavior is not exactly the same but its essence is the same, e.g., when a goods receipt is booked into the incorrect stock and only the assigned stock differs between the deviations. 
The third option provides the broadest insight into deviating behavior, but it requires domain knowledge for accurate grouping as well as sufficient data and the technical ability to aggregate deviations accordingly. If this approach is not feasible, one of the other options should be chosen.
%Regardless of the aggregation method, it is essential to understand the true essence of the deviation from a business perspective because similar deviating behaviors can stem from different causes and business scenarios, each having distinct implications that may affect the evaluation. When possible, these scenarios should be distinguished and analyzed separately. 

\mypar{Evaluate impact} In this step, the impact of the deviating behavior on the process and organization is evaluated to decide whether it is positive, negative, or neutral. This evaluation happens first from a direct perspective, reflecting on the immediate impact a deviation has, followed by a risk and opportunity perspective with impacts that do not materialize directly but only with a certain likelihood. Although both direct impacts and risk-and-opportunity-based assessments can draw on historical data, a forward-looking approach is especially important for the latter to identify a comprehensive set of possible scenarios. To assess the impact, four input factors are required:   
%In this step, we evaluate the general impact of the deviation based on its compliance, its outcome impact (i.e., what the goal), its performance impact (i.e., how the goal is achieved)...

\begin{compactitem} 
\item \textit{Compliance:} A deviation may violate compliance rules, which can be either internal, like best practices and internal policies, or external, like contracts and laws \cite{Dumas2015enabling,maggi2014predictive,andree2024amIallowed}. External rules are typically stricter than internal. %Violations of external compliance rules are generally seen as more severe than internal ones.

\item \textit{Outcome:} A deviation might impact the goal fulfillment of the process, i.e., the outcome to be accomplished. To assess this impact on outcome, it is essential to consider the role and centrality of the affected parts of the process \cite{outmazgin2013exploring,andree2023beyond}. For example, in a hospital, the goal is to provide the correct treatment to patients. Skipping the crucial ``check symptoms'' step would significantly jeopardize this goal.

\item \textit{Performance:} Beyond the outcome itself, deviations can also influence how an outcome is achieved, typically assessed by the process's performance in terms of cost, profit, time, and quality \cite{Dumas2015enabling,outmazgin2013exploring}. Even a single deviation can already have a performance impact that spans beyond its own case, affecting other cases within or even outside the process \cite{outmazgin2023fromautomatic}.

\item \textit{Standardization:} Standardization involves a deviation's impact on the complexity and predictability of a process \cite{knoblich2025conceptualizing}. Typically, deviations increase variance, e.g., by introducing alternative paths. Unlike performance, which focuses on execution, standardization affects BPM activities such as planning, monitoring, or training, where effort increases when standardization decreases.
\end{compactitem} 

%\mypar{Decision} 
To assign a deviation a positive, neutral or negative impact, all four factors have to be considered together since they potentially influence each other. Furthermore, the direct impact and the risk and opportunity impact should be combined to arrive at an all-encompassing impression of the impact.
%An example for a direct impact could be that if a purchase approval is skipped in a procurement process, this at first saves time. However, there is the chance for the risk-based impact of additional time and effort for returns in case management objects to the purchase in retrospect. 
If this evaluation step results in a neutral impact, the deviation is assigned \textit{neutral deviation}. For these deviations, the action recommendation again is to ignore them since they do not affect the process, and hence no reaction to them would be worthwhile. For deviations with a positive or negative impact, the framework continues with an assessment of possible reactions.
\update{M2, R2.3}{The skip of "Receive Payment" in Fig.~\ref{fig:DeviationExample} decreases processing time (positive performance impact) but violates the process intention and internal rules (negative outcome and compliance impact). As the latter outweighs the former, the deviation has negative impact overall.}

\mypar{Evaluate possible reaction} 
If a deviation shows a positive or negative impact, the next and last step identifies and evaluates potential reactions to proactively address %its cause and impact for 
future occurrences. For positive deviations, this involves measures to foster its adoption in the process. For negative deviations, preventive reactions should be considered. This step involves assessing the economic viability of the reaction, considering two input factors:

\begin{compactitem}
\item \textit{Reaction effectiveness:} The effectiveness of a reaction to a deviation can be defined by its ability to secure or amplify the impact of positive deviations or to reduce the impact of negative deviations. This depends both on the choice of measure as well as its support among process stakeholders. In particular, reaction effectiveness is determined by the analyst's capability to clearly identify the deviation's origin to look beyond symptoms and address the true reason behind it~\cite{alter2014theory,outmazgin2023fromautomatic}.

\item \textit{Reaction cost:} Reactions to a deviation entail certain cost. These can be one-off such as a re-iteration of a compliance training, or ongoing such as additional checking activities. Furthermore, there can be costs associated with change management. 
\end{compactitem} 

%\mypar{Decision} 
If the effectiveness of a chosen reaction is less than or equal to its cost, the deviation is classified as a \textit{reaction-inefficient deviation} and should be ignored, regardless of its impact. Conversely, if the effectiveness exceeds the cost, the deviation is labeled as either a \textit{positive deviation} or a \textit{negative deviation} depending on its impact from the prior step. For positive deviations, the framework recommends adopting them into the standard process, which might involve actions such as training resources in the new approach. For negative deviations, it advises implementing preventive measures, such as compliance awareness initiatives, IT system adjustments, or enhanced monitoring. 
\update{M2, R2.3}{In Fig.~\ref{fig:DeviationExample}, the execution of "Receive Payment" can be enforced by the underlying software.}

\update{M2, M4, M5, M6, R2.1, R2.2, R2.4, R3.3, R3.4, R3.6}{For details on the framework as well as options to adapt it to specific use cases, we provide extensive supplementary materials, which describe and exemplify input factors and output categories, and include a guideline to specify and prioritize input factors.\footnote{\url{https://doi.org/10.6084/m9.figshare.29064068}}}
%This marks the conclusion of the framework.

\section{Evaluation} \label{sec:eva}
We evaluated our framework to assess whether it effectively provides guidance to process analysts when analyzing concrete deviations in practice. In particular, we wanted to collect feedback about its completeness and correctness \update{R1.4}{(P1)}, as well as its usefulness for orientation by structuring the desirability assessment  \update{R1.4}{(P2)}.
In the following, we first present the setup of our evaluation before reporting on the evaluation results.

\subsection{Evaluation Setup}
We designed the evaluation as a deviation desirability assessment task that we conducted as semi-structured interviews. For those, we invited eight process analysis practitioners who had not taken part in the framework development interviews, but fulfilled the same selection criteria of process analysis experience (on avg. 14.6 years) and expertise in the respective process domain (see below, on avg. 16.1 years).
In contrast to the framework development interviews, we pursued a confirmatory instead of an exploratory goal for the evaluation and adjusted the assessment task accordingly. In particular, we extended the variety and detail of deviations and let experts assess them directly with our framework. 
We designed deviation examples from four different processes, namely \textit{intracompany replenishment in retail}, \textit{make-to-stock in discrete manufacturing}, \textit{invoice-to-cash in wholesale}, and \textit{procure-to-receipt in custom manufacturing}.\footnote{We provide an overview of participants and the interview guideline in the supplementary material online: \url{https://doi.org/10.6084/m9.figshare.28560650}}  
%For each process, we individually interviewed two participants specialized in the respective domain to compare their answers for consistency.

The interviews started with a presentation of the framework and its purpose. % to clarify questions the participants might have as well as collect feedback on the first impression. 
Then, we outlined a fictional company and its process to contextualize the task. For each process, we introduced the corresponding model, four deviating traces, and supplemental case context describing aspects such as potential exceptional conditions, technical details or attributes of the deviating case to make it more specific.
%provided four deviations, depicted using a BPMN model with an overlaying deviating trace, a textual description, and additional case context information. 
%These deviations included each output category of our framework at least once for each process. 
%These deviations were intentionally varied and kept open to make it possible to select each output category of our framework at least once across all interviews.  
The framework accepts any conformance deviation, but we opted for high-level deviation patterns \cite{grohs2024beyond} added as a text to each deviation for concise yet concrete descriptions. These patterns are control-flow-centric. While the framework is not limited to this perspective, we chose it to reduce cognitive load, as multi-perspective deviations add complexity and may be less familiar to participants. This focus ensures that evaluation results reflect the framework’s effectiveness without confounding factors from multi-perspective models.
As an example, consider Fig.~\ref{fig:DeviationExample} again which  was also used in the evaluation.

The task for the experts was then to assess each deviation's desirability by applying our framework.
For the depicted example, an expert could quickly render it a true deviation since no evidence points at data quality issues, model errors or exceptional conditions. A negative impact becomes apparent from the direct effect of missing pay as well as risks concerning the cash flow and the correctness of financial statements. A possible reaction could be to technically tie the clearing of invoices to a payment receipt as a preventive control measure, most likely being reaction-efficient, rendering the example a \textit{negative deviation}.
\update{M3, R1.1, R2.5}{Since the high-level deviation data provided in our laboratory evaluation set up requires interpretative assumptions that render classification outcomes incomparable, we did not evaluate whether participants arrive at a certain outcome, but focused on their feedback about the framework itself and their flow throughout its application.}
We asked the experts to state whenever they felt the need to diverge from the framework, how they wanted to diverge, and why. These potential divergences could then indicate a gap in the framework or an inadequately defined assessment step, allowing us to draw specific conclusions on our evaluation criteria of correctness and completeness.
%to get insights on correctness and completeness.
Concluding the interviews, we collected feedback on their experiences with the framework, particularly regarding the usefulness for orientation but also any other benefits and challenges they perceived.
\footnote{\update{M2, M5, R2.1, R2.3, R2.4, R3.3, R3.4, R3.6}{A detailed demonstration of an assessment can be found online: \url{https://doi.org/10.6084/m9.figshare.29097455}. This demonstration can serve as a blueprint for other assessments.}}

\subsection{Evaluation Results}
We coded the feedback along the framework's structure and the evaluation criteria  using a template approach. Overall, the framework was received very positively by the process analysts. Opportunities for improvement were mentioned only sparingly. Table~\ref{tab:comments} provides an excerpt of the experts' comments, on which we elaborate in the following.

\renewcommand{\arraystretch}{1.2}

\begin{table} [h]
    \vspace{-1em}
    \caption{Excerpt of expert comments from the evaluation interviews}
    \centering
    \begin{tabular}{|m{0.02\textwidth} | p{0.42\textwidth} | p{0.42\textwidth}|} 
        \hline
        &
        \textbf{Positive feedback} & 
        \textbf{Opportunities for improvement} \\
        \hline
        \multirow{2}{*}{\rotatebox[origin=c]{90}{\parbox[c]{4.3cm}{\centering \textbf{Correctness \& completeness}}}} & 
        \begin{minipage}[t]{\linewidth}
        \begin{itemize}[leftmargin=*,topsep=0pt] 
            \item "I have done this kind of analysis for years. It was exactly the approach that you have presented here" (P13)
            \item “As always, you could switch the order of some things but you can also keep it this way […]. It is correct to do it this way” (P15) 
            \item “The exemplary approach is universally applicable, because you have this approach everywhere” (P12)
        \end{itemize}
        \end{minipage}
        &
        \begin{minipage}[t]{\linewidth}
        \begin{itemize}[leftmargin=*,topsep=0pt]
            \item “You always have tolerance levels [for neutral deviations], where a certain deviation from the standard is expected” (P14)
            \item “Another approach [for impact evaluation] could be: to create an end-to-end process view.“ (P10)
            \item “At the end, you draw a comparison over all deviations, which one is the most relevant […], what should I deal with first?” (P11)
        \end{itemize}
        \end{minipage} \\
        \hline
        \multirow{2}{*}{\rotatebox[origin=c]{90}{\parbox[c]{6.8cm}{\centering \textbf{Usefulness for orientation}}}} & 
        \begin{minipage}[t]{\linewidth}
        \begin{itemize}[leftmargin=*,topsep=0pt]
            \item “This foundational classification, what do we perceive as a deviation, that is something that would be of relevance for a process analysis tool” (P11)
            \item “So for someone who is new to process mining, for them it is definitely great. So they get an impression: when is a deviation really a deviation […] what do I need to pay attention to?” (P10) 
            \item “I noticed that with your examples it definitely fits. I always know where I am” (P12) 
            \item “The false positives are very critical […]. If you do not filter well, the analysis costs much more” (P13)
        \end{itemize}
        \end{minipage}
        &
        \begin{minipage}[t]{\linewidth}
        \begin{itemize}[leftmargin=*,topsep=0pt]
            \item “You need deeper knowledge, it is not for high level senior management. Other than that, it is definitely useable and applicable to the issue.” (P12)
            \item “In general, it offers orientation. However, the framework is generalized and therefore abstract […]. That is always an act of balance.” (P15)
        \end{itemize}
        \end{minipage} \\
        \hline
    \end{tabular}
    \label{tab:comments}
    \vspace{-2em}
\end{table}

\mypar{\update{R1.4}{P1:} Correctness \& completeness} 
No expert reported any missing input factors or output categories or a misalignment of those with the evaluation steps while analyzing the deviation examples.
They especially highlighted the real-world fidelity of the framework's structure and the resulting natural analysis flow.
\update{M3, R1.1, R2.5, R3.5}{Mostly, the experts converged on similar assessments.}
Furthermore, they emphasized the framework's universal applicability across different processes.

The experts also pointed out some opportunities for improvement. One expert advised to incorporate a notion of tolerance levels for neutral deviations since many processes are subject to expected natural variance. Since our framework deliberately remains abstract in assigning a final impact in step four, we consider this covered by the framework's flexibility, but it might make sense to point this out in an accompanying documentation. 
Furthermore, some experts suggested explicitly incorporating an organizational end-to-end perspective, since a deviation might be an uncontrollable exception or a neutral deviation for the process at hand, but this might not be the case when considered from the perspective of the full value chain. As the goal of our framework is to support process analysts, who normally operate on the process level, this is out of scope for our framework, but might be an interesting opportunity for future work. Lastly, one expert proposed an additional step to the framework: They explained that after a possible reaction has been assessed for one distinct deviation, they would progress towards a prioritization among different deviations to decide which one to address first. Again, we consider this out of scope for our framework, which is meant to classify one deviation behavior at a time, but we agree that it is an interesting outlook on the next operative steps happening after the framework's termination. 

\mypar{\update{R1.4}{P2:} Usefulness for orientation} We observed that the experts underwent a learning phase regarding the use of the framework over the course of assessing the four deviations. While analyzing the first deviation example, most experts started by brainstorming on the deviation's desirability and then mapped this to the framework in retrospect, enhancing their assessment if they forgot steps or input factors. Reflecting on this, several experts underlined the importance of the checks in the first part of the framework which can easily be overlooked in ad-hoc assessments, leading to unnecessary analysis effort. 
For later deviations, they then started to directly apply the framework once they got more experienced in its usage.
Overall, they stated that one of the main benefits of the orientation offered by the framework lies in structuring every aspect that analysts need to consider during the assessment, which can be especially helpful for novices. The experts wished to have this reflected in process analysis tools.

Regarding opportunities for improvement, two experts noted that the framework's abstract level assumes certain domain knowledge to fill in factors, which might only be available to analysts in closer contact with the process as opposed to upper management. Since these are our main target group, we consider this an acceptable prerequisite.

\section{Discussion \& Conclusion}\label{sec:conlusion}

In this paper, we develop a procedural framework for assessing the desirability of process deviations on the basis of existing literature as well as interviews with process analysis practitioners and researchers. In five steps, our framework guides process analysts in deriving actionable insights from past deviations to proactively handle future occurrences. %Specifically, it identifies which assessment factors should be considered in what order to categorize deviations into one of seven mutually exclusive desirability categories. 
The evaluation demonstrates the framework's ability to streamline analysts' otherwise ad-hoc assessments into a comprehensive, yet concise deviation analysis.

\mypar{Future Research} We designed our framework as a generic, process-agnostic guide, which can be customized to specific needs. For instance, input factors could be refined into relevant key metrics or weights could be added to them to align with organizational priorities. Thus, framework variants could be tailored to specific process analysis and improvement objectives. 
\update{R2.7}{Also, we want to investigate the crucial step \textit{Evaluate impact} in more detail. This step encompasses multiple dimensions between which trade-offs might occur, e.g., in case a deviation saves time but has negative compliance impact. These kinds of trade-offs are typically discussed with the management of the process. We want to explore options to structure this critical step and support decision-making by both analysts and managers.}
From a procedural point of view, it could also be interesting to explore which other structural formats the framework could be integrated into. Future studies could, for example, examine whether formats such as iterative approaches building on a repetition of the sequential structure could enable progressive shifts from high-level to more detailed analysis to make the framework flexibly scalable for analysis scenarios with a high number of deviations.
%For example, it could be examined if our streamlined, sequential arrangement of assessment steps could serve as the foundation of a cyclic or iterative approach that allows to progressively detail out analysis if needed.
Additionally, as one expert in the evaluation pointed out possible next steps after the framework's termination, it could also be investigated how to bring the framework's insights into action and which additional needs such as the suggested subsequent prioritization of deviations could arise. 

%Methodological Limitations
\mypar{Methodological Limitations} Although our goal was to develop a process-agnostic framework, we were able to test it on only a limited number of processes. To mitigate this, we selected procurement as a representative process frequently used in process mining research for our framework development and evaluated four additional processes for broader insights. In the evaluation, we received positive feedback about the framework's universal applicability. 
\update{M3, R1.1, R2.5, R3.5}{Due to the framework's general applicability, we prioritized breadth over of depth for the evaluation, which came with the limitation of not being able to test it in a naturalistic setting with real-world data.
Further, we were not able to quantitatively evaluate if users assign correct output categories as this assignment is highly subjective, especially in laboratory settings. Consequently, any evaluation based on the correctness of answers was infeasible. Nevertheless, the experts mostly converged on similar assessments. We acknowledge this limitation but also do not think that correct assessments are the main contribution of our framework. Rather, we believe that the correctness, completeness \update{R1.4}{(P1)}, and usefulness \update{R1.4}{(P2)} of the framework itself are the main contributions, on which we received positive feedback.}
To increase transparency regarding the generalizability of our qualitative findings, we provide supplementary materials on all development and evaluation steps, incorporated a diverse participant pool from various job roles and process domains, and systematically processed findings through transcripts and structured coding. However, further application across different processes and domains is necessary to fully validate the framework’s versatility.
%Future studies could also complement our qualitative findings with a quantitative perspective. We adopted a qualitative approach because this allows us to obtain broad insights into the framework’s benefits, challenges, and their underlying reasoning. 
Now that our initial results suggest advantages such as real-world fidelity and reduced analysis effort, the next step could involve experimental surveys or case studies in natural settings to examine specific interest in greater detail. 

%Conceptual Limitations
\mypar{Application Challenges} The framework depends on specific data prerequisites, which may not always be fully met in practice. However, access to sufficient and high-quality data is crucial for accurately evaluating process deviations, such as uncovering issues in event logs or linking deviations to contextual conditions.
\update{R3.2}{Thus, in settings with limited contextual data, the framework can only be applied after data quality issues have been addressed.}
Future research could explore how to increase data availability, potentially leveraging Large Language Models and Internet of Things to also access less prominent data sources such as e-mail communications or sensor data.
\update{M4, M5, M6, R2.1, R3.3}{In addition, the framework  mandates domain-specific adaptions for concrete use cases as it is generic by choice to ensure universal applicability. Although this reduces detail, we believe that a generic framework is necessary to connect all aspects of deviation desirability. To address this, a potentially required domain-specific adaption can build on our supplementary materials, which provide guidelines for the prioritization of certain factors as well as a descriptions of input factors and output categories. Further, we provide a blueprint demonstration of a framework application.}
\update{M5, R1.2, R2.1, R3.4}{Another challenge is that the framework may impose a high cognitive load on users since it encompasses many qualitative input factors. However, we believe that this is due to the task at hand, which itself is highly complex and involves multiple stakeholders. The provided supplementary materials can help to reduce the cognitive load of users, which can already familiarize themselves with the framework, as exemplified in our evaluation where they gained experience for later analyses. 
In this light, future work could investigate tool support to further reduce cognitive load and required domain knowledge. Interest in such a tool support was also voiced during the evaluation experiments.}
\update{R1.3,R3.5}{Also, there might be ambiguity within output categories, meaning that, e.g., a deviation might be an exception in some cases whereas it is a negative deviation in other cases. Similarly, data availability limitations can lead to uncertainty, especially when trying to allocate a deviation to the log or model in situations with little evidence for their quality \cite[p.236]{CarmonaCCfund}. In future works, we aim to investigate how the framework allows for context-dependent assessments or hybrid categorization under uncertainty, e.g., by assigning fuzzy likelihoods to output categories. However, we already see the framework and its collection of input factors as an asset to communicate such issues transparently and identify uncertainties as a potential risk in interpreting results.}
Last,  behaviorally identical deviations can stem from different root causes, each carrying distinct desirability implications. In its current state, our framework deliberately abstracts from root causes, as they become relevant at a more granular level where factor complexity and ambiguity increase significantly, posing a threat to our goal of providing clear orientation. %We want to investigate whether there are ways of incorporating a root cause perspective inside certain evaluation steps to enhance analytical depth while maintaining a simple analysis structure, for example with iterative setups where parts of the framework are repeated to progressively detail out analysis when needed.
We want to investigate whether root causes can be incorporated without excessive complexity, building on empirical studies to establish links between root causes and deviation desirability, and assessing how such insights can enhance our framework.

\bibliographystyle{splncs04}
\bibliography{sample}

\end{document}